\documentclass[aps,pra,twocolumn,nopacs,superscriptaddress,groupedaddress]{revtex4}

\usepackage{amsmath}
\usepackage[dvips]{graphicx}
\usepackage{epsfig}
\usepackage{tabularx}
\usepackage{color}
\usepackage[pdfpagemode=UseNone,colorlinks=true,linkcolor=blue,citecolor=blue]{hyperref}

\usepackage{soul}
\usepackage{gensymb}

\begin{document}


\title[]{Simultaneous cooling of all six degrees of freedom of an optically levitated nanoparticle by elliptic coherent scattering}

\author{A. Pontin}
\email{a.pontin@ucl.ac.uk}
\affiliation{Department of Physics and Astronomy, University College London, Gower Street, London WC1E 6BT, United Kingdom}

\author{H. Fu}%
\affiliation{Department of Physics and Astronomy, University College London, Gower Street, London WC1E 6BT, United Kingdom}%

\author{M. Toro\v{s}}%
\affiliation{School of Physics and Astronomy, University of Glasgow, Glasgow, G12 8QQ, United Kingdom}%

\author{T. S. Monteiro}%
\affiliation{Department of Physics and Astronomy, University College London, Gower Street, London WC1E 6BT, United Kingdom}%

\author{P. F. Barker}
 \email{p.barker@ucl.ac.uk}
\affiliation{Department of Physics and Astronomy, University College London, Gower Street, London WC1E 6BT, United Kingdom}%

\begin{abstract}
We report on strong cooling and orientational control of all translational and angular degrees of freedom of a nanoparticle levitated in an optical trap in high vacuum.  The motional cooling and control of all six degrees of freedom of a nanoparticle levitated by an optical tweezer is accomplished using coherent elliptic scattering within a high finesse optical cavity. Translational temperatures in the 100 $\mu$K range were reached while temperatures as low as 5 mK were attained in the librational degrees of freedom. This work represents an important milestone in controlling all observable degrees of freedom of a levitated particle and opens up future applications in quantum science and the study of single isolated nanoparticles.
\end{abstract}

\maketitle

The field of optomechanics has, over the last twenty years, demonstrated exquisite control over the mechanical motion of oscillators. Some of these systems, whose masses range from the atomic to the kilogram scale, can be cooled to near their zero point energy allowing the creation of non-classical states of motion\cite{seis} and the study of quantum mechanics in new macroscopic regimes\cite{silanpaa}. The ability to engineer their coupling to light, as well as to other interactions such as magnetic \cite{gavartin2012hybrid} and electric fields \cite{regal}, or even gravity \cite{middlemiss}, allows them to be utilized as highly efficient frequency converters and for sensitive, compact and configurable sensors for a range of applications.

More recently, the ability to levitate nano and micromechanical oscillators using optical, electrical or magnetic fields has progressed to the point where their oscillatory motion can also be cooled to the ground state using both feedback cooling ~\cite{Magrini2021} and via coherent scattering within a high finesse optical cavity\cite{Delic2020science, marin2021}.  Coherent scattering (CS)~\cite{novotny_CS_2019,Delic_CS_2019,tania_CS_2020,tania_CS_2021}, a technique originally developed in the context of cavity QED~\cite{Vuletic2000} is solely driven by the light directly scattered by the particle itself into a high finesse optical cavity.

Levitated systems in high vacuum are extremely well isolated from the environment and have relatively few degrees of freedom to control. A particularly unique feature of these systems, unlike other optomechanical systems, are their rotational degrees of freedom. The rotational motion is particularly interesting as it offers an operational framework which goes beyond standard optomechanics as it is not limited to the dynamics of a quantum harmonic oscillator~\cite{kuhn2017,Stickler_2018,tennis,marko_precession}, and can be driven to rotate at very high speed~\cite{tongcanli_nanorotor_2018,novotny_nanorotor2020}. Angular DoFs can also be confined to a given orientation, recovering the usual harmonic dynamics. In this case, the motion is usually referred to as librational. Rotational dynamics of levitated nanoparticles represents more than a different arena for novel quantum dynamics since a  practical application of this work is in materials and bio-science where the orientation of nanoparticles in the absence of a substrate is required for diffraction imaging \cite{miao1999extending}. While initial promising work has been performed in a liquid \cite{yuan} such imaging requires orientational/librational damping with fluctuations in the $\mu$rad range which cannot be achieved in this environment. This can however be achieved by strong cooling of all DoFs.

Currently, most experimental efforts in controlling the librations have been focused on cooling the motion by active feedback methods~\cite{novotny_libration_fbk2021,tongcangli_5D}, either linear or quadratic, with a lowest effective temperature reached for the librational motion of $0.24$\,K~\cite{novotny_libration_fbk2021}. A more  promising approach relies on extending the coherent scattering framework to nonspherical nanoparticles. This has been explored theoretically~\cite{elliptic_CS_prl,elliptic_CS_pra} and it has been shown that with an asymmetric top nanoparticle it should be possible to simultaneously enter the quantum regime for both the CoM and librational motion. In this configuration the optical tweezer field needs to be elliptically polarized and the dynamics of two orthogonally polarized cavity modes must be considered.

\begin{figure}
\includegraphics[width=8.6cm]{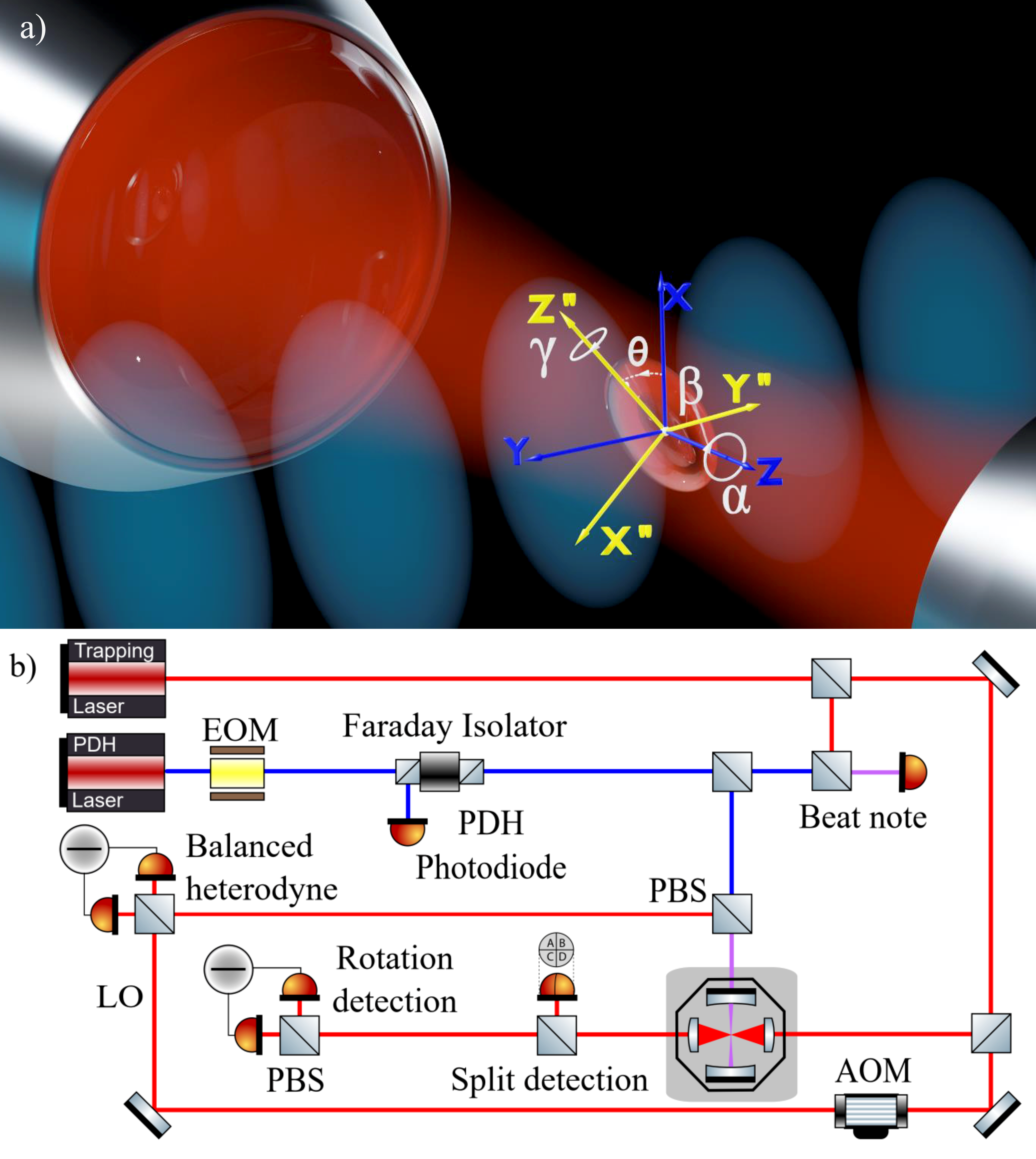}
\caption{Overview of the experiment. a) An optical tweezer propagating along the $z$ axis and polarized along $z''$ traps a prolate ellipsoidal silica nanoparticle. The transformation of coordinates from the particle body frame to the lab frame is performed by three consecutive rotations parametrized by three Euler angles $\Omega=(\alpha, \beta, \gamma)$ in the $z$-$y'$-$z''$ convention. The optical cavity symmetry axis is aligned with $y$. b) Simplified layout of the experiment\,\cite{supp}. A weak probe beam is used to lock the cavity by implementing a PDH scheme. A beam from a second laser illuminates a single aspheric lens (NA=$0.77$) to obtain the tweezer trap. The two lasers are offset locked one FSR apart ($\sim 12.3$\,GHz). Light collected by a condenser lens (NA=$0.77$) is used to measure the particle motion. An heterodyne detection allows to analyze light transmitted by the cavity.}
\label{fig1}
\vspace{-0.5cm}
\end{figure}

Here, we experimentally demonstrate the cooling of all six rototranslational DoFs of an ellipsoidal silica nanoparticle exploiting a CS approach. We measure an effective temperature for two librational DoFs on the order of $5$\,mK exploiting sideband resolved cooling. Furthermore, the two center of mass DoFs in the tweezer polarization plane are cooled to a few hundreds of $\mu$K, thus approaching the quantum regime. At the same time, the CoM motion along the tweezer propagation direction and the third remaining librational motion are also cooled ($\sim K$).

Our optical trap is composed of a single aspheric lens, nominal NA$=0.77$, with a symmetric collection lens and is illuminated by a $\lambda\simeq 1064$\,nm optical field with a filling factor slightly smaller that $1$. A silica nanoparticle is initially captured at atmospheric pressure and with tweezer field linearly polarized along the $x$ axis (see Fig.\,\ref{fig1} for the reference frame definition). The monolithic tweezer assembly is mounted on a 3-axes translational stage and placed at the centre of a high finesse Fabry-Perot cavity of length $L_{\text{cav}}=12.23\pm0.02$\,mm and which has a finesse of $\mathcal{F} \simeq 31000$ (half linewidth $\kappa/2\pi=198\pm1$\,kHz, input rate $\kappa_{\text{in}}=81\pm1$\,kHz) and a waist of $60\,\mu$m. A second Nd:YAG laser is phase modulated to implement a Pound-Drever-Hall (PDH) locking scheme. The two lasers are offset locked one free spectral range (FSR) away (FSR=$c/2L_{\text{cav}}= 12.26\pm0.02$\,GHz), with the PDH beam kept at resonance and the detuning of the trapping field which can be finely controlled. The light collected after the tweezer assembly is distributed on three balanced detectors, two exploiting a "\textit{split}" scheme where the beam is divided in two sections by a D-mirror (along orthogonal directions), while for the last one the beam is divided in two equal parts exploiting a polarizing beam splitter (this will be referred to as "\textit{PBS}" detection in the following). Light coherently scattered by the particle and driving the cavity is analyzed by a balanced heterodyne detection after being transmitted by the input mirror. The local oscillator is shifted by $f_{\text{LO}}=2$\,MHz from the trapping light. Both lasers are mode-matched to the cavity to better than $92\%$. The effective tweezer (TW) parameters are a power at the waist of $235$\,mW and a waist parallel and perpendicular to the polarization of $w_{\parallel}=831$\,nm and $w_{\perp}=771$\,nm.

\begin{figure*}[t]
\includegraphics[width=1\textwidth]{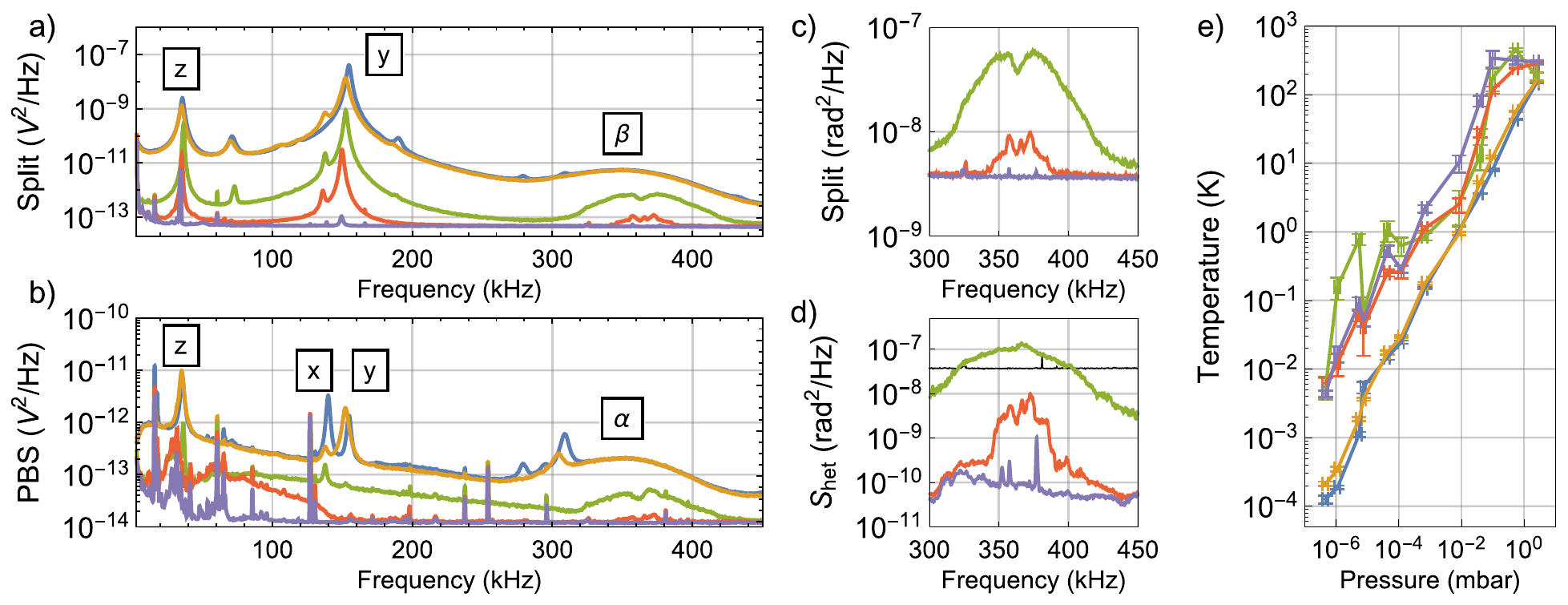}
\caption{Measured PSDs and effective temperatures for the nanoparticle DoFs at different pressures. \textbf{Panel a)}: PDSs from the \textit{split} detection monitoring the CoM $y$ motion which also detects librational motion of $\beta$. \textbf{Panel b)}: PSDs from the \textit{PBS} detection which measures librations of $\alpha$ along with a small contribution from $x$,$y$,$z$.
\textbf{Panel c)}: \textit{Split} PSDs around the librational spectral features; At the lowest pressure (violet) the small librational peaks can be seen at $357$\,kHz and $377$\,kHz.
\textbf{Panel d)}: In the lower pressure range explored the $PBS$ detection is not sensitive enough (black line shows its sensitivity) and we rely on the heterodyne detection of the cavity output. Here, the peak at $352$\,kHz and the features composing the noise floor are due to laser frequency noise. Librational peaks are the same as in Panel c).
\textbf{Panel e)}: Effective temperatures as a function of pressure; $x$, $y$, $z$, $\alpha$ and $\beta$ are shown in blue, yellow, green, red and violet respectively. The lowest temperatures obtained are $126\pm17\,\mu$K, $212\pm16\,\mu$K, $6\pm2$\,mK, $6.3\pm1.4$\,mK and $4.4\pm0.5$\,mK in the same order. In panels a)-d) spectra in blue, yellow, green, red and violet are taken at a pressure of $2.5$ (no cavity), $2.5$, $4\times10^{-2}$, $6.9\times10^{-4}$ and $1.1\times10^{-6}$\,mbar respectively.}
\label{fig2}
\end{figure*}

The TW field is propagating along the \textit{z}-axis, initially linearly polarized along \textit{x}, while the cavity axis lies along the y direction. The TW polarization state can be controlled both in terms of ellipticity and orientation in the \textit{x}-\textit{y} plane, by a combination of quarter and half waveplates, so that the general state can be defined as $\textbf{e}_{\text{t}}=\mathcal{R}(\theta,\textbf{e}_{\text{z}})\, \textbf{e}_{\text{p}}$ where $\mathcal{R}(\theta,\textbf{e}_{\text{z}})$ is a 3D rotation by an angle $\theta$ about $\textbf{e}_{\text{z}}$ and $\textbf{e}_{\text{p}}= \cos(\psi)\, \textbf{e}_{\text{x}} + i\, \sin(\psi)\,\textbf{e}_{\text{y}}$ defines the ellipticity ($\psi \in [0,\pi/4]$). Finally, transformation from the body to lab frame is performed with Euler angles $\Omega=(\alpha, \beta, \gamma)$ in the common $z$-$y'$-$z''$ convention and the diagonal susceptibility tensor $\chi=$diag$(\chi_1,\chi_2,\chi_3)$, in the body frame, has the following hierarchy $\chi_1< \chi_2< \chi_3$. In this setting a prolate-like particle will align, in steady state, to the TW polarization. For example, for a linear TW $\langle\beta\rangle=\pi/2$ and $\langle\alpha\rangle=\theta$ so that fluctuations of $\alpha$ are confined to the polarization plane while $\beta$ represents a rotation outside it.

One of the typical difficulties in levitated optomechanics consists in assessing with good accuracy the particle parameters. For a non spherical one, significant insight can be obtained from the asymmetry of the gas damping for the CoM motion. The experimental damping ratios in absence of the cavity are $\gamma_{\text{y}}/\gamma_{\text{x}}=1.200\pm0.008$ and $\gamma_{\text{z}}/\gamma_{\text{x}}=1.21\pm0.01$ which indicate a likely cylindrical symmetry but still allowing for a small asymmetry in the y-z plane (shape asymmetry up to $\sim2$\, nm). We determine the likely geometry exploiting the latter measurements and the five trap frequencies (see~\cite{supp} for details). We consider an ellipsoidal silica nanoparticle with $R_1=83.7\pm0.5$\,nm, $R_2=84.2\pm0.9$\,nm and $R_3=109\pm2$\,nm, density $\rho=1850$ and relative dielectric constant $\epsilon_{\text{r}}=1.98\pm0.07$~\cite{g2paper}. The method in Ref.~\cite{Cavalleri2010} also allows us to estimate the gas damping for the rotational DoF, we find for all DoFs $\gamma_i/2\pi\simeq P\,(533,751,758,646,702,694) [$Hz/mbar$]$ where $i=(x,y,z,\alpha,\beta,\gamma)$ and $P$ is the pressure (in mbar).

The dynamical equations for the nanoparticle motion under the combined action of the tweezer and cavity fields can be derived from the following Hamiltonian:

\begin{equation}
\label{eq1}
H=H_{\text{free}}+H_{\text{opt}}-\hbar \Delta (a^{\dag}a+b^{\dag}b)
\end{equation}

\noindent where $H_{\text{free}}$ is the free nanoparticle Hamiltonian, $H_{\text{opt}}$ is the total optical potential and the last term represent the contribution from the two orthogonally polarized TEM$_{\text{00}}$ cavity modes. In the Rayleigh approximation the optical term can be written as $H_{\text{opt}}=-(\epsilon_0 V_{\text{p}} /4) \mathbf{E}^*(\mathbf{r})\cdot\chi(\Omega)\mathbf{E}(\mathbf{r})$~\cite{elliptic_CS_pra} where $V_{\text{p}}$ is the particle volume and $\mathbf{E}(\mathbf{r})$ is the total electric field at the particle position $\mathbf{r}$. Non conservative contributions due to radiation pressure force and torque must also be included as well as the stochastic driving associated with thermal noise~\cite{rot_thermal_noise_1,rot_thermal_noise_2,rot_thermal_noise_3}.

The total electric field is given by the sum of the TW field $\mathbf{E}_{\text{TW}}(\mathbf{r})$ and the cavity fields $\mathbf{E}_{\text{cav}}(\mathbf{r})$ populated by light scattered by the particle. Thus, the optical potential will have three contributions (i) a term $\propto\mathbf{E}^*_{\text{TW}}(\mathbf{r})\cdot \chi(\Omega) \mathbf{E}_{\text{TW}}(\mathbf{r})$ which represents the TW trapping potential (ii) a term $\propto\mathbf{E}^*_{\text{cav}}(\mathbf{r})\cdot \chi(\Omega) \mathbf{E}_{\text{cav}}(\mathbf{r})$ which is the usual intensity coupled dispersive interaction~\cite{giacomo2015,giacomo2016,Kiesel14180}, and (iii) a term $\propto\mathbf{E}^*_{\text{TW}}(\mathbf{r})\cdot \chi(\Omega) \mathbf{E}_{\text{cav}}(\mathbf{r})+ h.c.$ which is the CS interaction term that would not be present if the TW field would not have been resonant with the cavity.

The CS Hamiltonian of Eq.~\ref{eq1} provides a path to 6D cooling, however, cooling rates cannot be simultaneously optimal for all DoFs. Cooling of the translational modes in the TW polarization plane is optimized at a cavity node while all other DoFs are more efficiently cooled at an antinode. Typically, another difficulty emerges from the hierarchy of the trap frequencies which imposes limits for the choice of the cavity detuning with direct consequences on cooling rates. The theory of elliptic CS is presented elsewhere~\cite{elliptic_CS_pra} with a complete analysis in the deep trapping regime, i.e. when all six DoFs are stably trapped and the orientation of the particle is fixed. In this regime the equations of motion can be safely linearized. However, when $\gamma$ is not confined the intrinsic strong nonlinearity of the rotational dynamics emerges dominantly with $\alpha$ and $\beta$ nonlinearly coupled to each other through the motion of $\gamma$~\cite{tongcangli_5D,Robicheaux2019_PRA}.

The operational parameters, chosen to strike a necessary compromise for the cooling efficiencies, are a detuning of $\Delta/2\pi=-362$\,kHz\,$\simeq(\omega_{\alpha}+\omega_{\beta})/2$ and a particle mean position in the cavity standing wave of $y_o\sim0.155\,\lambda$ (antinode at $y_o=0$). This choice takes full advantage of the relatively small frequency separation between $\omega_{\alpha}$ and $\omega_{\beta}$ and the fact that these frequencies are not deeply sideband resolved so that a significant cooling rate is still present for the CoM. The mean position $y_o$ is kept closer to the node to reduce the contribution of frequency noise. Finally, we used an ellipticity of the TW field of $\psi=14.5^{\circ}\pm1^{\circ}$, which is the largest value we could set to avoid spinning the particle in the entire pressure range explored.

We show in Fig.\,\ref{fig2}\,a)-b) experimental power spectral densities (PSD) at different pressures for the \textit{split} detection orthogonal to the TW polarization fast axis and for the \textit{PBS} detection. With a simple extension to what has already been modeled~\cite{Robicheaux2019_PRA,supp} it is possible to show that to lowest order the \textit{split} detection is sensitive to the $\beta$ motion while the \textit{PBS} detection captures the motion of $\alpha$. At high pressure, and in absence of the cavity (blue trace), $\gamma$ undergoes free diffusion. As has already been shown~\cite{tongcangli_5D,Robicheaux2019_PRA}, this leads to a strong nonlinear coupling between $\alpha$ and $\beta$ so that one has to redefine the dynamical normal modes which will have frequencies of $\omega_{\pm}=\sqrt{(\omega_{\alpha}^2+ \omega_{\beta}^2+\omega_{c}^2\pm Q)/2}$. Here, $Q=\sqrt{4 \omega_{\beta}^2\omega_c^2+(\omega_c^2+\omega_{\alpha}^2- \omega_{\beta}^2)^2}$ and $\omega_c\simeq(I_3/I_1)(\dot{\gamma}-\dot{\alpha}\,\beta)$ which implies that the frequency difference is time dependent. On short timescales $\omega_{\pm}$ appear as two narrow spectral peaks, but over longer times, over which the intrinsic thermal evolution of $\omega_c$ has been exhaustively sampled, they manifest as a broad Gaussian peak. In the spectra shown in Fig.~\ref{fig2}, this transition is clearly visible since the total observation time $T_{\text{S}}$ per spectra is almost constant while the autocorrelation time goes from much smaller to much larger than $T_{\text{S}}$ as the pressure is reduced. Importantly, as $\alpha$ and $\beta$ are cooled, $\omega_c$ will be dominantly determined by the evolution of $\gamma$ so that the $\omega_{\pm}$ frequency difference should provide direct insight in the behavior of $\gamma$. 

The calibrated low pressure spectra for $\beta$ and $\alpha$ are highlighted in Fig.~\ref{fig2}\,c)-d) respectively. The \textit{PBS} detection is, however, significantly less sensitive compared to the \textit{split} detection. This problem can be bypassed by setting the heterodyne detection to analyze the fluctuations of the cavity mode with polarization parallel to the TW polarization which ideally couples exclusively to $\alpha$. Spectra in Fig.~\ref{fig2}\,d) are obtained with the heterodyne but we also show the noise floor of the \textit{PBS} detection (black line). While the \textit{PBS} detection becomes less useful at low pressure, it is still necessary for the calibration of the $\alpha$ motion since it allows its measurement without the interaction with the cavity. Notice that to calibrate the spectra in Fig.~\ref{fig2}\,c)-d) knowledge of the moment of inertia is required, however, this is not the case for determining the effective temperature which relies only on the usual assumption of thermal equilibrium at high pressure ($\simeq$mbar) and in absence of the cavity.

By integrating the peaks area we measure the effective temperature of the three CoM DoFs and the two  tightly confined librational DoFs, i.e., $\alpha$ and $\beta$. These are shown in Fig.~\ref{fig2}\,e) as function of pressure. The lowest values measured are $126\pm17\,\mu$K and $212\pm16\,\mu$K for the \textit{x} and \textit{y} DoFs respectively while the librational DoFs reach $6.3\pm1.4$\,mK and $4.4\pm0.5$\,mK for $\alpha$ and $\beta$ respectively. The motion along the TW propagation remains somewhat hotter with an assigned effective temperature value of $6\pm2$\,mK. Analytical calculations based on the nominal parameters indicate that the recoil limit would occur at a lower pressure, in particular, for librational DoFs.

\begin{figure}[t]
\includegraphics[width=8.6cm]{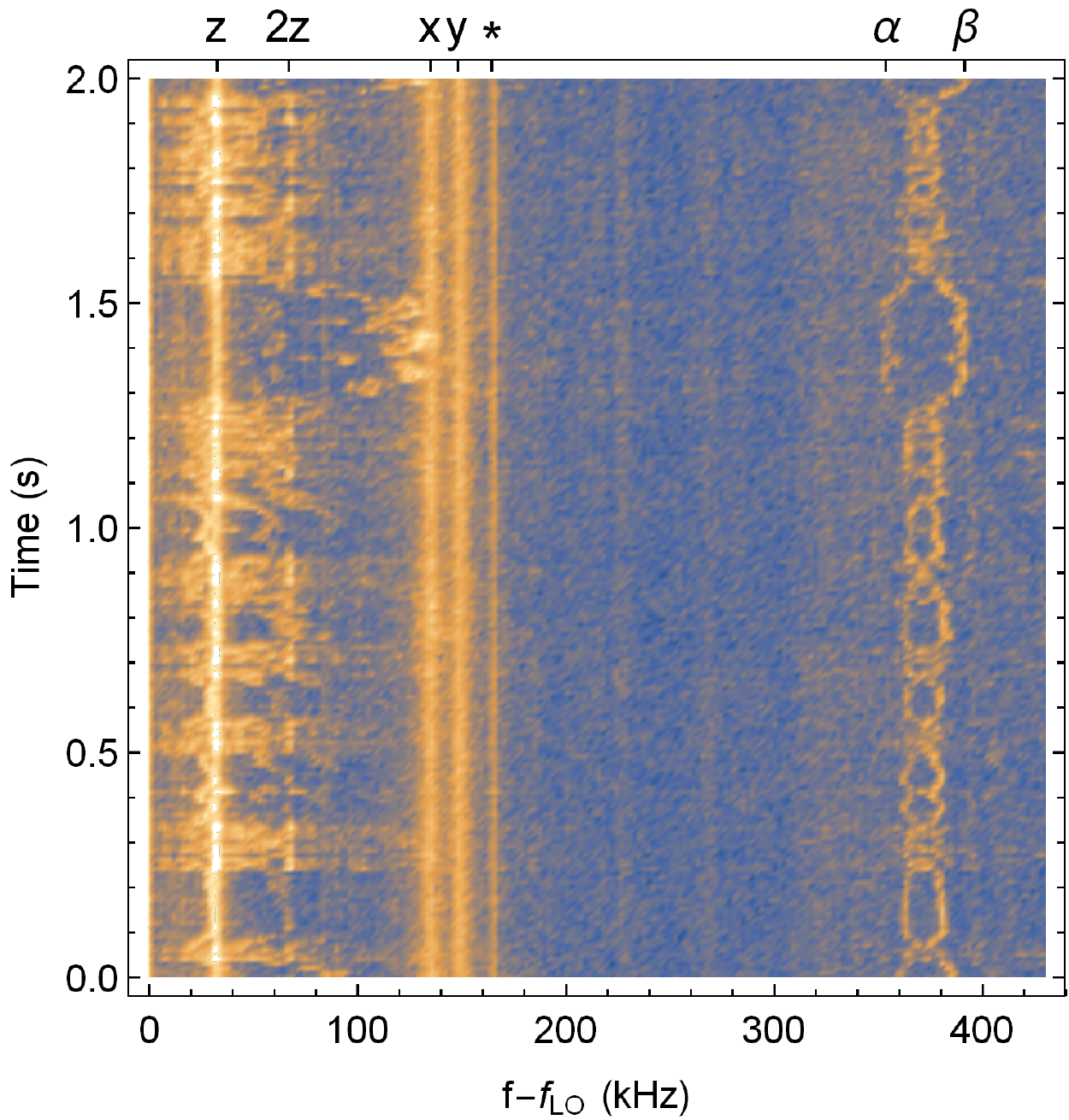}
\caption{Upper sideband spectrogram of the heterodyne detection at a pressure of $4.5\times10^{-5}$\,mbar. The cooled CoM resonances $x$, $y$, $z$ can be easily identified at $135$, $148$ and $33$\,kHz respectively. The time dependent frequencies for $\alpha$ and $\beta$ show the typical behavior associated with a diffusing $\gamma$ motion. In the frequency range $0$-$140$\,kHz, a fast evolving feature correlates extremely well with $\alpha$ and $\beta$ frequency difference and it is most likely due to the evolution of $\gamma$. The spectral feature marked with $(*)$ identifies an electronic spike.}
\label{fig3}
\end{figure}


A more intriguing aspect of the observed dynamics, however, can be found in the lower frequency band of the \textit{PBS} spectra in Fig.~\ref{fig2}~b). Initially, a $1/f^2$ PSD is clearly visible which transitions to a plateau-like structure at intermediate pressures and seemingly disappears at the lowest. More insight can be obtained from the heterodyne signal which offers a better sensitivity provided the CoM motion is sufficiently cold. We show in Fig.~\ref{fig3} a spectrogram of the upper sideband of the heterodyne covering $2$\,s of dynamics. The CoM peaks are clearly visible and stable in frequency, while the $\omega_{\pm}$ show the clear signature of diffusion in $\gamma$. Also present, is a low frequency peak highly time dependent which correlates exceptionally well with the frequency difference of $\omega_{\pm}$. This is particularly evident between $0.1$ and $0.2$\,s and right before $1.5$\,s. It also seems to couple to the z motion since a peak at $2\omega_z$ appears only when it is swinging around $\omega_z$. This might also explain the behaviour the \textit{z} effective temperature and its higher than expected value. Combining these observations is suggesting that this low frequency peak is actually related to the evolution of $\gamma$. For a cylindrically symmetric particle $\gamma$ should not be observable, however, this is no longer the case even for the small nominal difference of $\Delta_r=r_2-r_1=0.5$\,nm. For this value the analytical trapping frequency is calculated to be $\omega_{\gamma}/2\pi\simeq17$\,kHz which would increase to $33$\,kHz for the maximum $\Delta_r=1.7$\,nm allowed by the statistical uncertainty (one $\sigma$). The potential is, however, quite shallow with an estimated trap depth $\sim20$\,K. An intuitive picture of the evolution of $\gamma$ is that of a particle evolving in an infinite one-dimensional periodic potential driven by thermal noise. At high pressure, the potential is negligible so that $\gamma$ is freely diffusing. But there exists a critical pressure for which the evolution of $\gamma$ transitions from running to oscillating solutions. At intermediate pressures, both solutions coexist~\cite{Risken1989,tsallis_1,tsallis_2}; this is the regime captured in Fig.~\ref{fig3}.

\begin{figure}
\includegraphics[width=8.6cm]{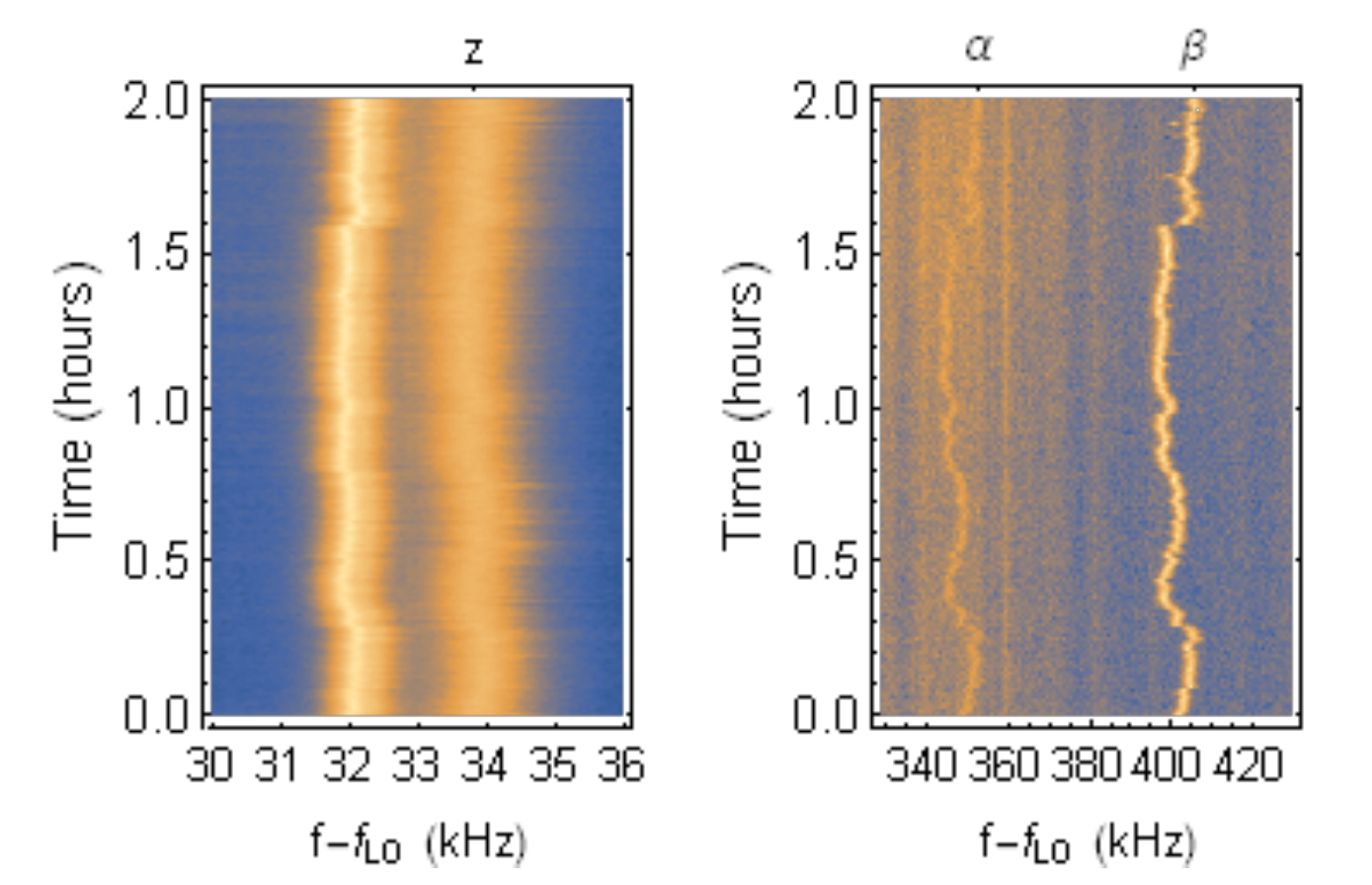}
\caption{Upper sideband spectrogram of the heterodyne detection at a pressure of $4.7\times10^{-7}$\,mbar, centred around $\omega_{\text{z}}$ (left) and $\omega_{\pm}$ (right). Here, the spectrograms are separated, since the color scale covers $40$\,dB for the first and only $10$\,dB for the second. The spectrograms cover $2$\,hours of observation over which $z$, $\alpha$ and $\beta$ are quite stable apart from small common mode drifts unlike the situation at higher pressures (e.g., see the green line in Fig.~\ref{fig2} c)) where a diffusive behaviour of $\gamma$ induces broad Gaussian peaks for $\alpha$ and $\beta$. This behavior over such long timescales ($\gg1/\gamma_{\alpha,\beta,\gamma}$) indicates that the orientation of the particle is fixed and the peak at $32$\,kHz is $\gamma$.}
\label{fig4}
\end{figure}

We show in Fig.~\ref{fig4} two spectrograms in a frequency bandwidth around $\omega_z$ and around $\omega_{\alpha,\beta}$ at a pressure of $4.7\times10^{-7}$\,mbar, this time covering $\simeq2$\,hours of observation. The spectral peak associated to $\gamma$ is, in this case, quite stable around $32$\,kHz and follows the same common mode slow drift that can be seen on all other DoF shown, suggesting that these are not related to dynamics but rather drifts in power or detuning. This situation is consistent with $\gamma$ being confined which would completely fix the orientation of the particle and imply that $\gamma$ is cooled as well by the interaction with the cavity to an effective temperature significantly smaller than the trap depth. The overall behaviour for the angular DoFs as the pressure is reduced is also reproduced relatively well in numerical simulations~\cite{supp}.  Nominally, the cavity mode that is analyzed with the heterodyne detection should not be coupled to $\gamma$ and should not carry information about its evolution. However, this is no longer the case if the orthogonal mode has a finite mean photon occupation. This can occur in presence of a small tilt of the TW propagation direction with respect to the cavity. Another possibility could be found in the TW true field distribution. Typically CS models relies on a Gaussian approximation for the TW beam which is clearly not accurate when high NA lenses are employed. In this case there is a small but nonvanishing polarization component along the TW direction. While this can be neglected when focusing on the CoM motion, it might have significant impact on the rotational dynamics. A better understanding will require further investigation, for example, by implementing a second heterodyne detection for the mode currently not analyzed.




In conclusion, we have demonstrated cooling of all the observable degrees of freedom of a levitated asymmetric nanoparticle via elliptic coherent scattering paving the way for future experiments that have been proposed for the creation and control of nanoclassical states of rotational motion. However, as optimal cooling for librational motion cannot be achieved at the same time as translation, careful optimisation of both cavity and particle parameters will be required to reach these goals in the future. As most levitated nanoparticles are non-spherical, protocols for the creation of macrosocpic center-of-mass superpositions will also require localisation of particle orientation as demonstrated here \cite{PhysRevLett.119.240401,PhysRevLett.127.023601}. Finally, our work represents the first time that an nanoscale object has been cooled and oriented in high vacuum. At the lowest pressures, the standard deviation of the trapped angular motion in the $\alpha$, $\beta$, $\gamma$ angles are 0.05, 0.04, 17 degrees respectively. The ability to fix the orientation of a nanoparticle in this way, free from interference from a substrate, offers the potential for single nanoparticle characterisation via coherent x-ray diffraction or electron diffraction, where the particle and source angle could be maintained or varied in a controlled way.

A.P. thanks Francesco Marin for useful discussion. The authors acknowledge funding from the EPSRC Grant No. EP/N031105/1. H.F. acknowledges the Engineering and Physical Sciences Research Council [grant number EP/L015242/1]. M.T. acknowledges funding by the Leverhulme Trust (RPG-2020-197).
\bibliographystyle{nicebib}
\bibliography{elliptic_CS_bib}

\end{document}